\documentclass[
aip,
amsmath,amssymb,reprint,showpacs]{revtex4-1}

\usepackage{graphicx}
\usepackage{epstopdf}
\usepackage[format=hang,singlelinecheck=0,font={sf,small},labelfont=bf]{subfig}

\newcommand{\FIG}[1]{Fig.~\ref{#1}}
\newcommand{\EQ}[1]{Eq.\,(\ref{#1})}
\newcommand{\TBL}[1]{Table~\ref{#1}}
\newcommand{\Poisson}{\nu_{P}}
\newcommand{\aGGG}{a_{s}} 

\newcommand{\prb}{Phys. Rev. B}
\newcommand{\pr}{Phys. Rev.}
\newcommand{\apl}{Appl. Phys. Lett.}	
\newcommand{\jmmm}{J. Magn. Magn. Mater.}
\newcommand{\jphysd}{J. Phys. D: Appl. Phys.}
\newcommand{\jap}{J. Appl. Phys.}
\newcommand{\natcomm}{Nat. Commun.}
\newcommand{\natphys}{Nat. Phys.}
\newcommand{\jcrgr}{J. Cryst. Growth}
\newcommand{\aplmat}{APL Mater.}
\newcommand{\ieeemagnlett}{IEEE Magn. Lett.}
\newcommand{\physstatsola}{Phys. Stat. Sol. (a)}

\begin{document}

\title{Sub-micrometer yttrium iron garnet LPE films with low ferromagnetic resonance losses}

\author{Carsten Dubs}
\author{Oleksii Surzhenko}
\author{Ralf Linke}
\affiliation{INNOVENT e.V., Technologieentwicklung, Pr\"ussingstr. 27B, 07745 Jena, Germany}
\author{Andreas Danilewsky}
\affiliation{Kristallographie, Albert-Ludwigs-Universit\"at Freiburg, Hermann-Herder-Str. 5, 79104 Freiburg, Germany}
\author{Uwe Br\"uckner}
\author{Jan Dellith}
\affiliation{Leibniz-Institut f\"ur Photonische Technologien (IPHT), Albert-Einstein-Str. 9, 07745 Jena, Germany}

\begin{abstract}
Using liquid phase epitaxy (LPE) technique (111) yttrium iron garnet (YIG) films with thicknesses of $\approx$100\,nm and surface roughnesses as low as 0.3\,nm have been grown as a basic material for spin-wave propagation experiments in microstructured waveguides. The continuously strained films exhibit nearly perfect crystallinity without significant mosaicity and with effective lattice misfits of $\Delta a^\perp / a_s \approx 10^{-4}$ and below. The film/substrate interface is extremely sharp without broad interdiffusion layer formation. All LPE films exhibit a nearly bulk-like saturation magnetization of (1800$\pm$20)\,Gs and an `easy cone' anisotropy type with extremely small in-plane coercive fields $<$0.2\,Oe. There is a rather weak in-plane magnetic anisotropy with a pronounced six-fold symmetry observed for saturation field $<$1.5\,Oe. No significant out-of-plane anisotropy is observed, but a weak dependence of the effective magnetization on the lattice misfit is detected. The narrowest ferromagnetic resonance linewidth is determined to be 1.4\,Oe @ 6.5\,GHz which is the lowest values reported so far for YIG films of 100\,nm thicknesses and below. The Gilbert damping coefficient for investigated LPE films is estimated to be close to $1\times 10^{-4}$.
\end{abstract}

\date{\today}

\pacs{81.15.Lm, 75.50.Gg, 76.50.+g}


\maketitle


\section{Introduction}
Magnonics is an increasingly growing new branch of spin-wave physics, specifically addressing the use of magnons for information transport and processing\cite{1,2,3,4}. Single crystalline yttrium iron garnet (YIG), which is a ferrimagnetic insulator with the smallest known magnetic relaxation parameter\cite{5}, appears to be a superior candidate for this purpose\cite{6,7,8}. As bulk or as thick film material, which is commonly grown by liquid phase epitaxy (LPE)\cite{9}, it has a very low damping coefficient and allows magnons to propagate over distances exceeding several centimeters\cite{6}. However YIG functional layers for practical magnonics should be nanometer-thin with extremely smooth surfaces in order to achieve optimum efficiency in data processing and dramatic reduction in energy consumption of sophisticated spin-wave devices. Therefore, high-quality thin and ultra-thin YIG films were grown using different growth techniques such as LPE, pulsed laser deposition (PLD) and rf-magnetron sputtering to investigate diverse spin-wave effects and to design YIG waveguides as well as nanostructures for spin wave excitation, manipulation and detection in prospective magnonic circuits.

From previous reports about sub-micrometer YIG films with thicknesses between 100 and 20\,nm\cite{10,11,12,13,14,15} available microwave and magnetic key parameters were taken and summarized in \TBL{tab:1}. Thus, ferromagnetic resonance (FMR) data were included which have been extracted from measurements of the absorption curves or absorption derivative curves versus sweeping magnetic in-plane field $H$ at a fixed frequency $f$ or vs. sweeping rf-exciting field $h_{rf}$ with an applied in-plane static magnetic bias field. The reported FMR linewidths $\Delta H$ and converted peak to peak linewidths $\Delta H_{p-p}$ of the field derivative values ($\Delta H=\sqrt[]{3}\Delta H_{p-p}$), which will be given during the further paper as full-width at half-maximum $\Delta H_\mathrm{FWHM}$, varied between 3\,Oe and 13\,Oe. The Gilbert damping coefficient $\alpha$ were found in the range from $2\times10^{-4}$ to $8\times10^{-4}$. Only the lowest given $\alpha$ value of $0.9\times 10^{-4}$ was obtained for a very short fit range of about 4\,GHz without any given data in the low frequency range below 10\,GHz \cite{13} and is therefore not really comparable with the other reported values. From this compilation it is obvious that neither $\Delta H_\mathrm{FWHM}$ nor $\alpha$ is significantly influenced by the YIG film thickness down to 20\,nm. The differences are probably resulted from additional ferromagnetic losses due to contributions of homogeneous and/or inhomogeneous broadening by microstructural imperfections or magnetic inhomogeneities.

\begin{table*}[bt!]
\caption{\label{tab:1}Key parameters reported for thin/ultrathin YIG films on (111) GGG substrates}
\centering
\begin{ruledtabular}
\begin{tabular}{ccccccccc}
Growth method  & Thick-  & RMS- & $4\pi M_s$\,\footnotemark[1] & $H_\mathrm{c}\,\footnotemark[1]$ & $\Delta H$\,\footnotemark[1] & $f_0$ & $\Delta H_0$\,\footnotemark[1] & $\alpha$  \\
(Reference) 	& ness	& roughness & &   & FWHM & & FWHM & $\times 10^{-4}$ \\
& (nm)	& (nm) & (kGs) & (Oe) & (Oe) & (GHz) & (Oe) &  \\
\hline
LPE\,\cite{10}			& 100 	& - 			& 1.81 		& -					& 3.0 		& 7 		& 1.6 					& 2.8 \\
LPE (this study) 		& 83--113  & 0.3--0.8 	& 1.78--1.82& $\leq$0.2			& 1.4--1.6	& 6.5 		& 0.5--0.7 				& 1.2--1.7 \\
PLD\,\footnotemark[2]\,\cite{11} & 79 	& 0.2 	& 1.72 		& $<2$				& 3.0 		& 10 		& 1.4 					& 2.2  \\
PLD\,\cite{12} 			& 23 	& - 			& 1.60 		& $<1$	& 3.5\,\footnotemark[3]	& 9.6 		& 3.5--7\,\footnotemark[3] 	& 2--4  \\
Sputtering\,\cite{13} 	& 22 	& 0.13 			& 1.78 		& 0.4	& 12\,\footnotemark[3] 	& 16.5 		& 6.4\,\footnotemark[3] 	& 0.9  \\
Sputtering\,\cite{14} 	& 20 	& 0.2 			& - 		& 0.4	& 13\,\footnotemark[3] 	& 9.7 		& 7\,\footnotemark[3]		& 8  \\
PLD\,\cite{15} 			& 20 	& 0.2--0.3 		& 2.10 		& 0.2	& 3.3\,\footnotemark[3] & 6 		& 2.4\,\footnotemark[3] 	& 2.3  \\
\end{tabular}
\end{ruledtabular}
\footnotetext[1]{Measurements at RT with the in-plane external magnetic field $H$}
\footnotetext[2]{YIG films grown on the (100) GGG substrates}
\footnotetext[3]{Peak-to-peak value $\Delta H_{p-p}$ of the derivative of FMR absorption transformed into $\Delta H_\mathrm{FWHM} = \Delta H_{p-p} \times \sqrt[]{3}$}
\end{table*}

In this report we present microstructural, magnetic and FMR properties of LPE-grown 100\,nm thin YIG and Lanthanum substituted (La:YIG) films with low ferromagnetic resonance losses. Film thicknesses were determined by X-ray reflectometry (XRR) and surface roughness by atomic force microscopy (AFM) measurements. Crystalline perfection and compositional homogeneity were investigated by high-resolution X-ray diffraction (HR-XRD) and X-ray photoelectron spectroscopy (XPS) as well as by secondary ion mass spectroscopy (SIMS). Static and dynamic (microwave) magnetic characterizations were carried out by vibrating sample magnetometry (VSM) and by Vector Network Analysis (VNA), respectively.

\section{Results}
\subsection{Microstructural properties}

\begin{figure}[tb!]
\begin{center}
\subfloat[]{\label{fig:XRR}\includegraphics[width=0.9\linewidth]{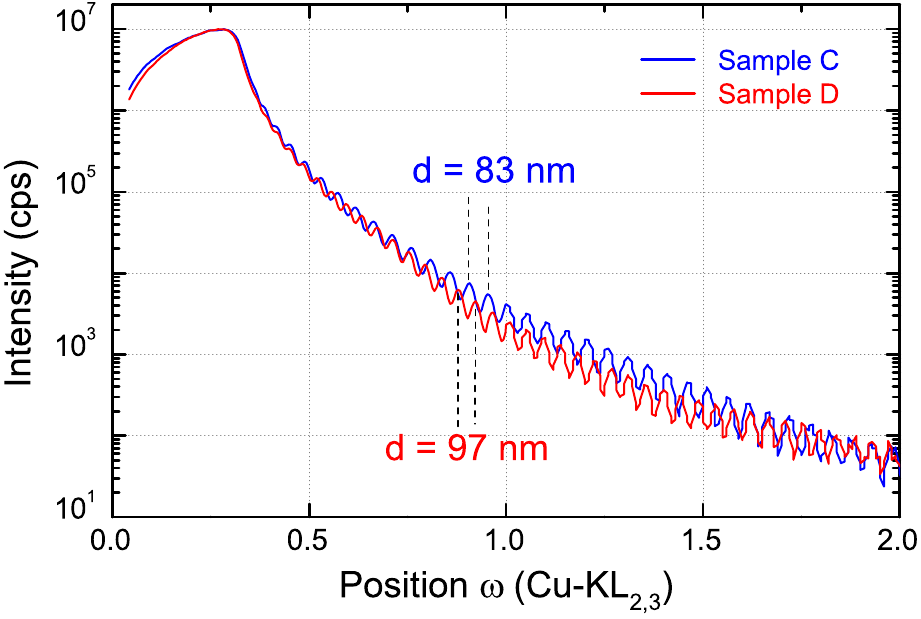}}\\
\subfloat[]{\label{fig:AFM}\includegraphics[width=0.9\linewidth]{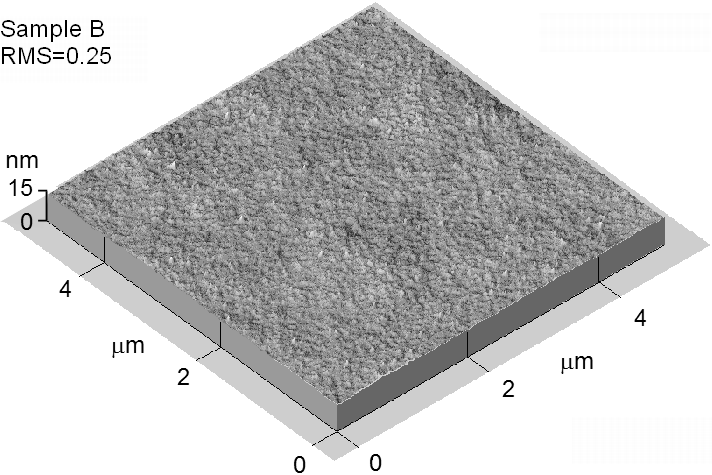}}
\end{center}
\caption{
(a) XRR plots of sub-micrometer-thick YIG LPE films.
(b) $5\times 5\,\mu m^2$ AFM surface topography of sample B with RMS roughness of 0.25\,nm.
}
\label{fig:1}
\end{figure}

Selected microstructural and magnetic properties of liquid phase epitaxial grown YIG (sample A-C) and La:YIG (sample D) films are given in \TBL{tab:2}. The consistent magnetic as well as microwave properties obtained for films deposited during different growth runs demonstrate a high reproducibility of the LPE growth technique. \FIG{fig:XRR} shows XRR plots of films with thicknesses of about 100\,nm which are smaller than the previously reported thinnest LPE YIG films\cite{17,18,19}. The smallest root-mean-square (RMS) surface roughness of about 0.25\,nm obtained for the sample B in \FIG{fig:AFM} is nearly comparable with epi-polished GGG substrate quality of $\approx$0.15 nm and with the best PLD and sputtered YIG films (see e.g. \TBL{tab:1}). Besides, films with slightly rougher surfaces (see \TBL{tab:2})) were obtained as a result of additional dendritic aftergrowth and/or due to plateau formation, so called ``mesas'', if any solution droplet adheres to the sample surface.

\begin{table*}[tb!]
\caption{\label{tab:2}YIG/La:YIG film properties grown on (111) GGG substrates by LPE technology}
\centering
\begin{ruledtabular}
\begin{tabular}{cccccccccc}
	& Thick-  & RMS- & Relative lattice & \multicolumn{2}{c}{VSM\,\footnotemark[1]} & \multicolumn{4}{c}{FMR\,\footnotemark[1]} \\
	\cline{5-6}\cline{7-10}
Sample& ness	& roughness & misfit $\Delta a^\perp/a_s$ & $4\pi M_s$ & $H_c$ & $4\pi M_\mathrm{eff}$ & $\Delta H_\mathrm{FWHM}$\,\footnotemark[2] & $\Delta H_0$ & $\alpha$\\
	& (nm)	& (nm) & $\times 10^{-4}$ & (kGs) & (Oe) & (kGs) & (Oe) & (Oe)  & $\times 10^{-4}$ \\
\hline
A 					& 113 		& 0.8 		& 4.7  	& 1.82 		& 0.10 		&	1.637	& 1.4 		& 0.5 		& 1.4 \\
B 					& 106 		& 0.3 		& 1.8	& 1.78 		& 0.20 		&	1.658	& 1.5 		& 0.7 		& 1.2 \\
C 					& 83  		& 0.6 		& 0.3  	& 1.82 		& 0.16 		&	1.672	& 1.4 		& 0.5 		& 1.6 \\
~~D\,\footnotemark[3] & 97  	& 0.8 		& 0.0 	& 1.78 		& 0.18 		&	1.712	& 1.6 		& 0.7 		& 1.7 \\
\hline
Accuracy			& $\pm 1$  	& $\pm 0.1$ & $\pm 0.3$ & $\pm 0.04$ & $\pm0.03$ 	& $\pm0.010$	& $\pm0.1$ 	& $\pm0.1$ 	& $\pm0.1$ \\
\end{tabular}
\end{ruledtabular}
\footnotetext[1]{VSM and FMR measurements at room temperature with applied in-plane magnetic field}
\footnotetext[2]{FMR linewidth value at frequency $f$=6.5\,GHz}
\footnotetext[3]{La:YIG LPE film}
\end{table*}

\begin{figure}[tb!]
\begin{center}
\subfloat[]{\label{fig:HR_XRD}\includegraphics[width=0.90\linewidth]{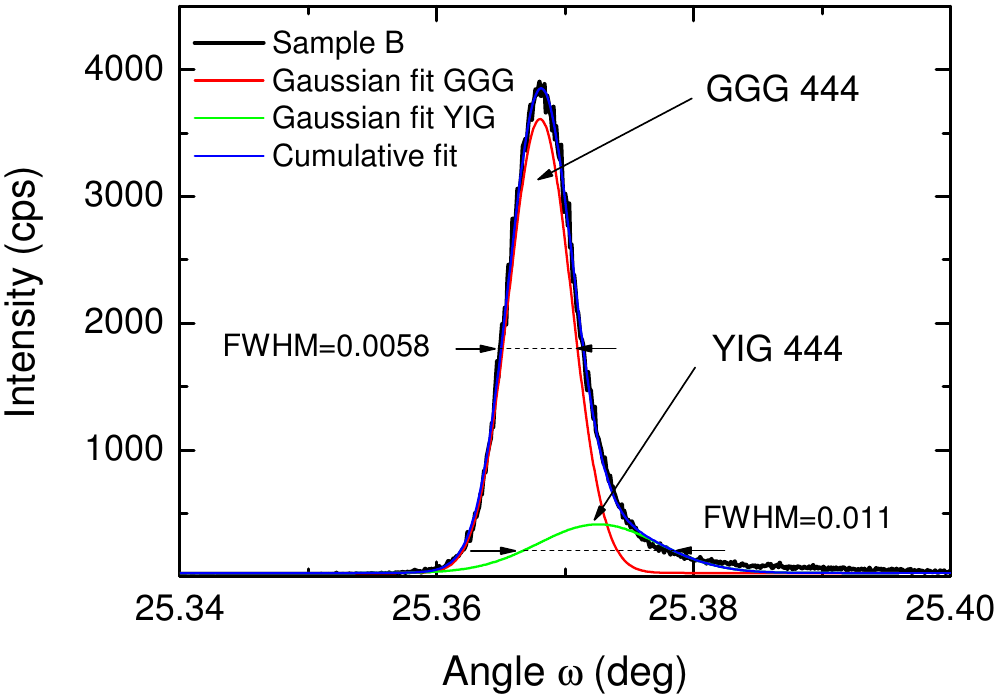}}\\
\subfloat[]{\label{fig:HR_RSM}\includegraphics[width=0.93\linewidth]{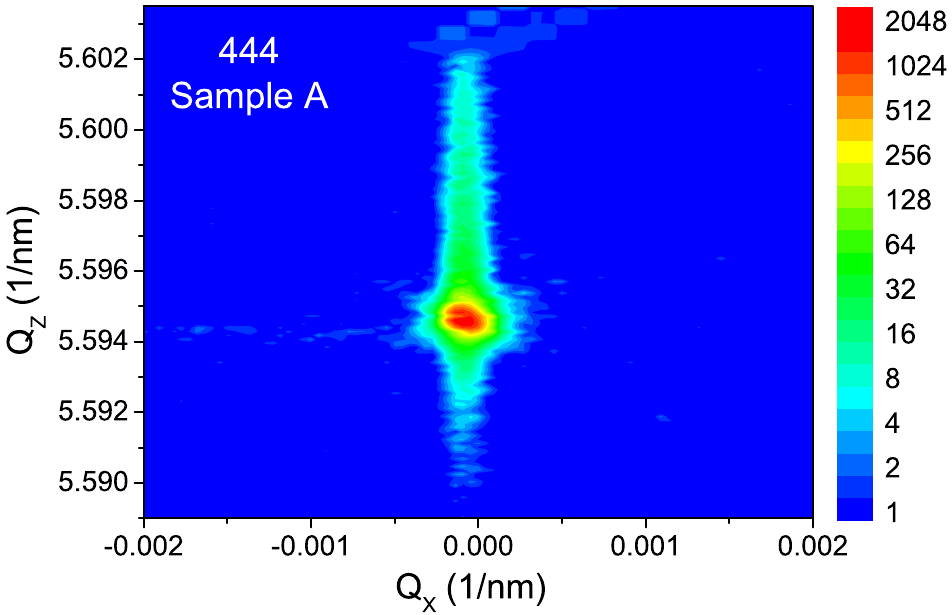}}
\end{center}
\caption{
(a) HR-XRD $\omega$-scan around substrate/film 444 Bragg reflection of sample B. By fitting procedures the YIG film peak has been extracted. 
(b) HR-RSM scans around substrate/film 444 reciprocal point reveal asymmetric diffracted intensities towards higher Q$_z$ values for sample A. 
}
\label{fig:XRay}
\end{figure}

HR-XRD studies of our thin epitaxial LPE films have been found to be difficult because of the nearly super-imposed diffraction pattern of YIG film and GGG substrate. Although the angle distances between film and substrate Bragg reflections were above the resolution limit of our HR-XRD equipment, the diffraction intensity of the film reflection was very low and results only in a broadening of the GGG Bragg reflection. \FIG{fig:HR_XRD} shows a $\omega$-scan (rocking curve) with a Gaussian-like fitted GGG substrate 444 reflection and a second fitted peak at the right shoulder which corresponds to the YIG 444 film reflection. This indicates a tensile stressed YIG film because of the smaller film lattice parameter compared to the commercially available Czochralski-grown GGG substrate ($\aGGG$=1.2382\,nm). For La:YIG films we observed a perfect pseudo-Voigt fitted substrate peak without any additional shoulder (not shown) which indicates a perfect lattice match between substrate and LPE film. This is in remarkable contrast to YIG films deposited by various gas phase techniques such as PLD and rf-sputtering \cite{20,21,11,12,13,14,15} on GGG substrates. For those films the YIG reflection has always been detected at considerably lower Bragg angles compared to the GGG substrate indicating a significant distortion of the cubic YIG garnet cell with significantly enlarged lattice parameters (compressive stress) \cite{20,23}.

The relative effective misfit $\Delta a^\perp / a_s = (\aGGG - a_\mathrm{YIG}^\perp)/\aGGG$ obtained from strained film lattice parameter in growth direction $a^\perp_\mathrm{YIG}$ and the substrate lattice parameter $\aGGG$ can be used as a measure for epitaxial induced in-plane tension or strain. Due to YIG Poisson's ratio of $\Poisson = 0.29$ pseudomorphously grown, fully strained YIG films with an ideal YIG$_{bulk}$ lattice parameter $a_\mathrm{YIG}=1.2375$\,nm\cite{24} should have a relative effective misfit of $\Delta a^\perp / a_s = + 11\times 10^{-4}$ (tensile stress). In the case of our sub-micrometer YIG films $\Delta a^\perp / a_s$ has been determined to be in the range between zero and $+5\times 10^{-4}$ (see \TBL{tab:2}) compared to PLD-grown YIG films with up to $\Delta a^\perp / a_s = - 100\times 10^{-4}$ (see e.g. Ref.\onlinecite{12}).  Hence, our LPE films are under tension but not to the extent which we expected for nominally pure YIG material without additional lattice expansion by lattice defects or impurities. To find the reason for this, high-resolution reciprocal space map (HR-RSM) and XPS investigations were performed. \FIG{fig:HR_RSM} shows a HR-RSM plot around the symmetrical 444 Bragg reflection with symmetrical diffracted intensity for the GGG substrate and asymmetric diffracted intensity toward higher scattering angles along $Q_z$ (2$\theta$-$\omega$-Scan) which we attribute to the YIG 444 film reflection. Broadening of the film reflection along $Q_z$ is due to the finite coherence lenght of the sub-micrometer thin film in growth direction and other broadening mechanisms as for example heterogeneous strain. The extension of the film reflection up to the substrate peak position suggests that the film is continuously strained due to an existing compositional and/or strain gradient. No peak broadening along the $Q_x$ direction ($\omega$-scan) indicates single crystalline perfection parallel to the film plane without significant mosaicity due to tilts of epitaxial regions with respect to one another.

\begin{figure}[tb!]
\begin{center}
\subfloat[]{\label{fig:XPS}\includegraphics[width=0.9\linewidth]{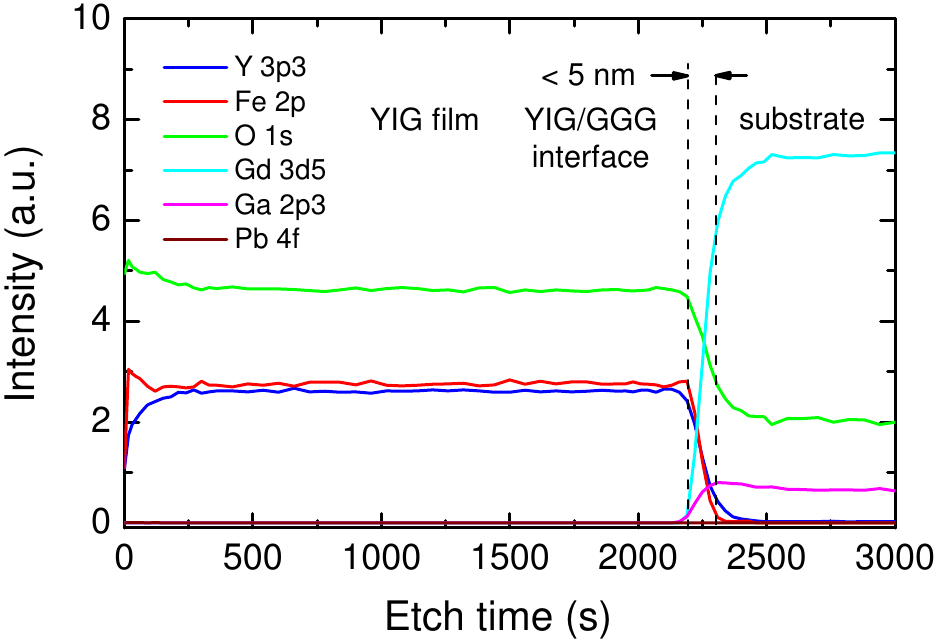}}\\
\subfloat[]{\label{fig:SIMS}\includegraphics[width=0.9\linewidth]{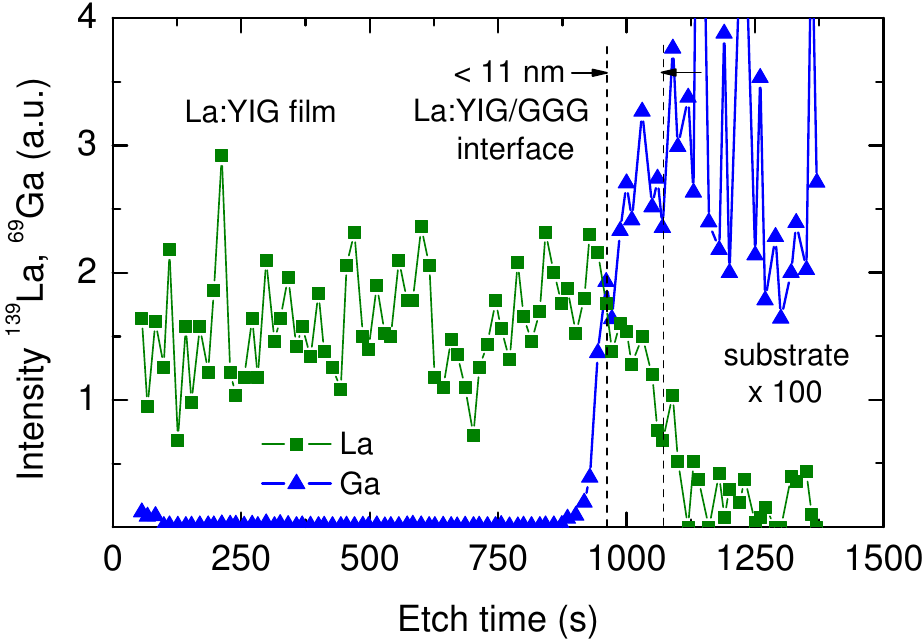}}
\end{center}
\caption{
(a) XPS depth profile of sample B reveals a very narrow interface between film and substrate. The Pb 4f signal could not be detected within the detection limit of about 0.1 at-\%. 
(b) SIMS depth profile analysis detects the $^{139}$La signal of the film as well as the $^{69}$Ga signal of the substrate (sample D) and their changes at the film/substrate interface.
}
\label{fig:Profiles}
\end{figure}

To evaluate the compositional homogeneity along the growth direction of the films and to detect expected impurities (e.g. Pb from solvent) depth profile analyses were carried out by XPS. \FIG{fig:XPS} shows a homogeneous distribution of the YIG matrix elements along the film growth direction and a sharp transition at the film/substrate interface. The obtained width of the transition layer for sample B is below 5\,nm. But the obtained depth profile consists of a convolution of the true concentration profile with the depth resolution of the XPS system under the concrete measuring conditions and should be narrower. Therefore, these profiles demonstrate that no broad interdiffusion layer is formed by element intermixing at the interface at an early state of epitaxial growth or by diffusion of substrate ions into the epitaxial layer and vice versa during the subsequent growth process.

Whereas XPS surface analysis of the very first atomic layers (not shown) gives a Pb content of about 0.2\,at-\%, no Pb signal could be observed during the depth profile analyses within the detection limit of 0.1\,at-\%\cite{25}. Therefore, it is assumed that the Pb signal corresponds to a surface contamination of condensed PbO vapor from high temperature solution and this contamination is completely removed by the first argon-ion etching step. For YIG films grown in La$_2$O$_3$ containing solution no La signal could be detected by XPS that give indicates that the La content must be below 0.5\,at-\%\cite{26}. In order to improve the detection capability additional qualitative SIMS measurements were carried out. Due to the resulting sputtering effect and by time-dependent detection of the sputtered sample ions one obtains depth profiles of the film elements as shown for $^{139}$La in \FIG{fig:SIMS}. Here, the counts of two separate measurements taken under identical measuring conditions at neighboring sample positions were added up in order to enhance the statistical significance. It is clearly visible that the lanthanum signal decreases at the film/substrate interface whereas substrate signals like $^{69.7}$Ga simultaneously increase.

\subsection{Static magnetic measurements}

The vibrating sample magnetometry  was used to measure the net magnetic moment $m$ of the YIG/GGG samples at room temperature. As a thickness of GGG substrates $\approx$5000 times exceeded these of the studied YIG films, a proper calculation of the YIG parameters required us (i) to extract the GGG contribution that linearly increased with the external field $H$ and (ii) to prefer the in-plane sample orientation that ensured considerably lower fields $H_{s}$ for the YIG films to attain the saturation.

\begin{figure}[tb!]
\begin{center}
\subfloat[]{\label{fig:VSM_Loop}\includegraphics[width=0.90\linewidth]{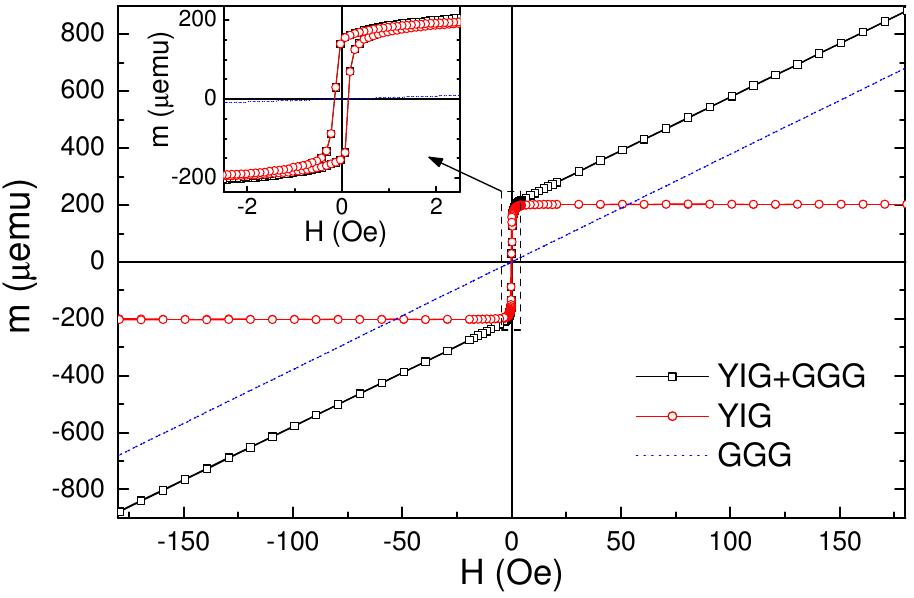}}\\
\subfloat[]{\label{fig:VSM_Anisotropy}\includegraphics[width=0.82\linewidth]{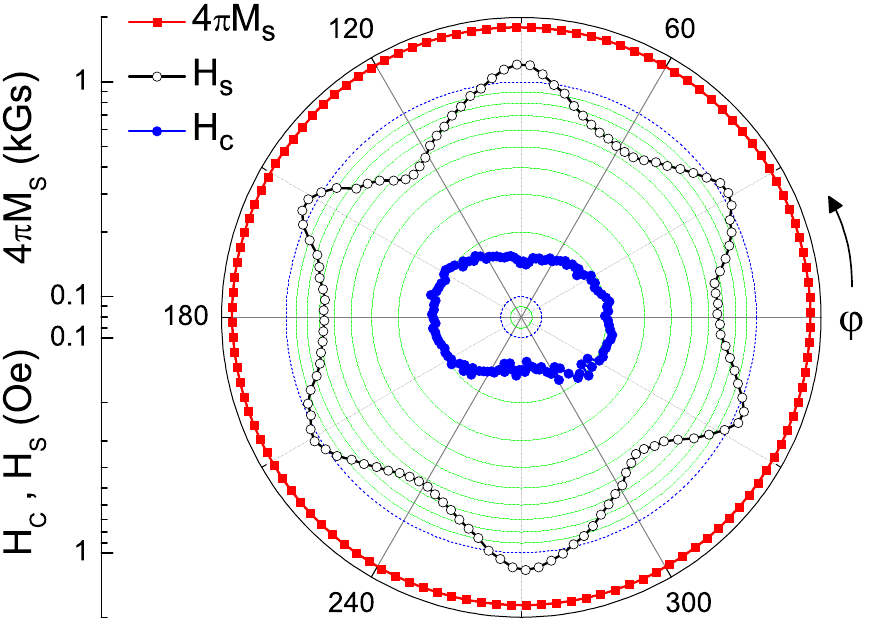}}
\end{center}
\caption{
(a) The net VSM magnetic moment $m$ of the sample D as well as its components induced by the YIG film and the GGG substrate vs the in-plane magnetic field $H$ parallel to the $\langle 110\rangle$ direction. 
(b) Azimuthal angle dependencies for the VSM loop parameters of sample D, i.e. a saturation magnetization $M_s$, a saturation field $H_s$ and a coercivity $H_{c}$. The $H_s$ six-fold symmetry with the mimima along the $\langle 112\rangle$ `easy axes' and the maxima along the $\langle 110\rangle$ `hard axes' indicates the cubic magnetocrystalline anisotropy.}
\label{fig:VSM}
\end{figure}

\FIG{fig:VSM_Loop} presents a typical dependence of the total magnetic moment $m$ vs the in-plane magnetic field $H$ and illustrates the method allowing us to separate the $m$' components produced by the YIG film and the GGG substrate. Being subsequently normalized to the film volume, the YIG component loops yield the following material parameters -- a saturation magnetization $M_s$, a coercivity $H_c$ and a saturation field $H_s$, i.e. the field (averaged over ascending $H_\uparrow$ and descending $H_\downarrow$ branches of hysteresis loops) where the YIG film magnetization approaches 0.9$\times M_s$. In order to estimate the in-plane anisotropy, we have repeated this procedure for the samples rotated around the $\langle 111\rangle$ axis perpendicular to film surfaces. \FIG{fig:VSM_Anisotropy} demonstrates such results as polar semi-log plots vs the azimuthal angle $\varphi$. A saturation magnetization $M_s$ in \FIG{fig:VSM_Anisotropy} seems independent of $\varphi$. The obtained $4\pi M_s$ values cluster around 1800\,Gs usually reported\cite{29} for bulk YIG single crystals. Within an experimental error (mostly defined by the YIG volume uncertainity of $\pm$2\%), the same is valid for the $4\pi M_s$ values in other LPE films listed in \TBL{tab:2}. The obtained coercivity ($H_c\leq0.2$\,Oe) in studied LPE films is among the best values reported for gas phase epitaxial films (see \TBL{tab:1}). No distinct influence of the crystallographic orientation on the $H_c$ values is also registered. In contrast, the azimuthal dependence of the saturation field $H_s$ obviously reveals the six-fold symmetry which matches the crystallographic symmetry of YIGs. The $H_s$ maxima coincide with the in-plane $\langle 110 \rangle$ projections of the hard magnetization axes, whereas the $H_s$ minima correspond to the $\langle 112 \rangle$ crystallographic directions. The $\langle 112 \rangle$ `easy axes' orientation suggests an `easy cone' anisotropy after Ubizskii\cite{30}. He has also demonstrated\cite{31} that relatively small in-plane magnetic fields lead to single-domain YIG films, although a deviation of magnetization vector from the film plane still remains due to finite values of the cubic anisotropy constants.
 
In conclusion, as the demagnetizing factor at the out-of-plane YIG film orientation is 1, the out-of-plane saturation field has to be close to the in-plane $4\pi M_s$ values. This fact is qualitatively confirmed by our out-of-plane measurements. Unfortunately, the GGG component of the total VSM signal at fields $H_\perp \approx 1.8$\,kOe is much larger than magnetic moments of YIG films with a thickness of $\approx 100$\,nm (see, for instance, \FIG{fig:VSM_Loop}) and, hence, a reasonable accuracy of $\pm$0.5\,\% at the GGG signal elimination inevitably results in too large errors for the YIG parameters. One may conclude that the out-of-plane configuration may provide reliable results when the ratio of YIG to GGG thickness exceeds, at least, $10^{-3}$.

\subsection{FMR absorption}

\begin{figure}[b!]
\begin{center}
\includegraphics[width=0.95\linewidth]{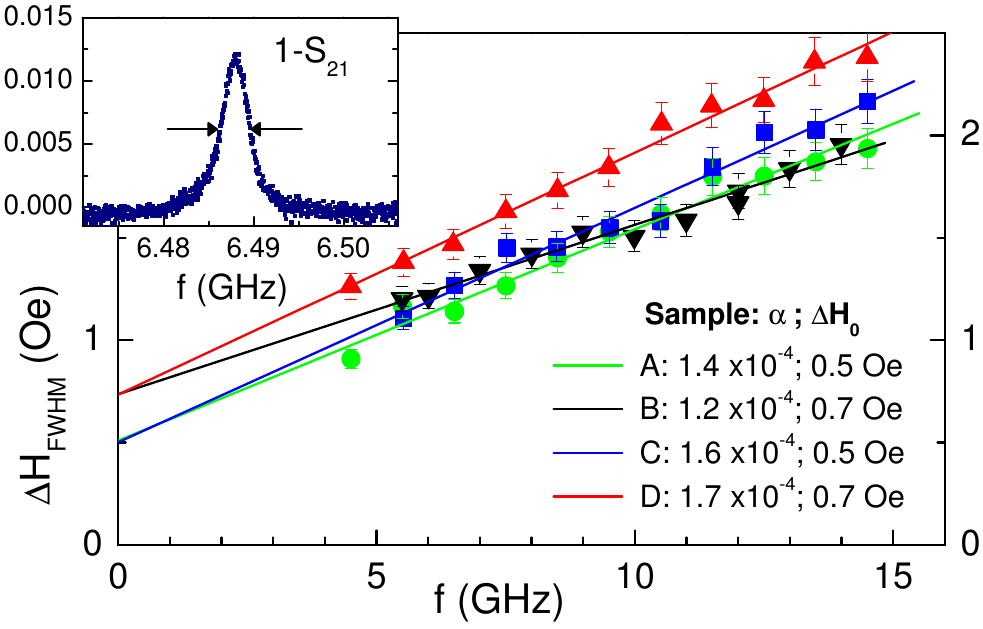}
\end{center}
\caption{Frequency dependence of FMR absorption linewidth $\Delta H_\mathrm{FWHM}$ for YIG LPE films A--D at various values of the in-plane magnetic field ($H\leq 5$\,kOe). Straight lines are linear fits that the Gilbert damping factors $\alpha$ are obtained from. Inset shows an example of FMR absorption spectrum measured for the sample A at $H=1.6$\,kOe.}
\label{fig:FMR_Linewidth}
\end{figure}

FMR absorption spectra for each of studied YIG films were recorded at several values ($H\leq 5$\,kOe) of the in-plane magnetic field. The inset in \FIG{fig:FMR_Linewidth} shows such a spectrum at $H=1.6$\,kOe that looks like the Lorentz function with a linewidth $\Delta f_\mathrm{FWHM}\approx4$\,MHz centered near the FMR frequency $f\approx 6.5$\,GHz. Since the FMR linewidth is mostly expressed in units of magnetic field, we, at first, used the centers $f$ of measured spectra and the corresponding in-plane fields $H$ to estimate the gyromagnetic ratio $\gamma$ and the effective magnetization $M_\mathrm{eff}$ in the Kittel formula\cite{Kittel48}

\begin{equation}   
f=\gamma\ \sqrt[]{H(H+4\pi M_\mathrm{eff})}.\\
\label{eq:Kittel}                                    
\end{equation}

\noindent Then, the best fitting pair of $\gamma$ and $M_\mathrm{eff}$ allowed us (i) to convert every frequency spectrum into the magnetic field scale, (ii) to fit rescaled spectra with the Lorentz function and (iii) to evaluate, thereby, the corresponding linewidth $\Delta H_\mathrm{FWHM}$. The selected results of the described procedure -- namely, $4\pi M_\mathrm{eff}$ and $\Delta H_\mathrm{FWHM}$ at the reference frequency $f=6.5$\,GHz -- are listed in \TBL{tab:2}, while the whole summary of the obtained $\Delta H_\mathrm{FWHM}$ values is presented in \FIG{fig:FMR_Linewidth} vs the FMR frequency. The plots in \FIG{fig:FMR_Linewidth} are known\cite{10,11,12,13,14} to provide data about the Gilbert damping coefficient $\alpha$ and the inhomogeneous contribution $\Delta H_0$ to the FMR linewidth that are mutually related by
\begin{equation}
\Delta H_\mathrm{FWHM}=\Delta H_0+\frac{2\alpha f}{\gamma}
\label{eq:Damping}     
\end{equation}

As the FMR performance of thin YIG films strongly depends on the working frequency of future magnonic applications, we have included various quality parameters in \TBL{tab:2}, \textit{viz.} i) the Gilbert damping coefficient $\alpha$ which is mostly responsible for the FMR losses at high magnetic fields ($H\gg 4\pi M_\mathrm{eff}$), ii) the inhomogeneous contribution $\Delta H_0$ that dominates at small fields ($H\ll 4\pi M_\mathrm{eff}$) as well as iii) the FMR linewidth at the reference frequency $f$=6.5\,GHz which approximately corresponds to the case $H\approx 4\pi M_\mathrm{eff}$. The latter is estimated down to $\Delta H_\mathrm{FWHM}$=1.4\,Oe that is to our knowledge the narrowest value reported so far for YIG films with a thickness of about 100\,nm and smaller. The Gilbert damping coefficients are estimated to be close to $\alpha\approx 1\times 10^{-4}$ which is comparable to the best values reported so far (compare with \TBL{tab:1}). The zero frequency term $\Delta H_0$ is found almost the same for all YIG films including the La substituted one. The obtained value $\Delta H_0\approx 0.5-0.7$\,Oe appears as well appreciably lower than that for gas phase epitaxial films (see \TBL{tab:1}).
 
\begin{figure}[tb!]
\begin{center}
\includegraphics[width=0.95\linewidth]{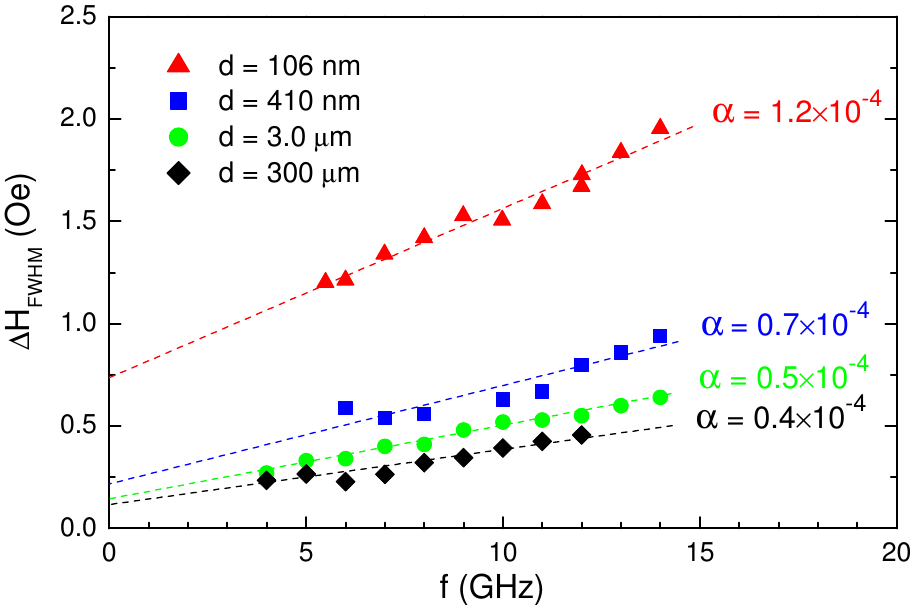}
\end{center}
\caption{Frequency dependencies of the FMR linewidth $\Delta H_\mathrm{FWHM}$ for YIG LPE films of various thickness $d$ and the YIG sphere with diameter $d=300\,\mu m$. The Gilbert damping factors $\alpha$ are calculated from slopes of the best linear fits according to \EQ{eq:Damping}.}
\label{fig:FMR_Width_vs_Thickness}
\end{figure}

In summary, optimized LPE growth and post-processing conditions improve FMR linewidths and Gilbert damping coefficients (compare this study and Ref.\onlinecite{32} with Ref.\onlinecite{10}). However, the improved values are still far from these in bulk YIGs and relatively thick YIG films (see \FIG{fig:FMR_Width_vs_Thickness}) due to the decreasing volume to interface ratio in sub-micrometer films. For example, imperfections at the film interface of thin films should have a stronger influence on the magnetic losses in contrast to the dominating volume properties of perfect thick films. It requires us to undertake further attempts to minimize the FMR performance deterioration with a decrease of the YIG film thickness. These attempts will be focused on avoiding the most probable sources of FMR losses such as contributions due to homogeneous broadening (interface roughness, homogeneously distributed defects and impurities) and inhomogeneous broadening (geometric and magnetic mosaicity, single surface defects) and, thus, on approaching the ``target'' parameters of $\Delta H_\mathrm{FWHM}=0.3$\,Oe at 6.5 GHz and $\alpha=0.4\times 10^{-4}$ reported by \citet{28} for bulk discs made of single YIG crystals.

\section{Outlook and conclusions}
Besides the efforts to avoid growth defects as well as interface roughness and to reduce impurity incorporation during the LPE deposition process further high-resolution investigations are necessary to gain more insight into the YIG microstructure and to identify the properties which play an essential role for its FMR performance. Therefore, in future studies we will carry out HR-RSM scans with asymmetrical reflections to determine in-plane and axial strain, respectively, the Time-of-Flight (ToF) SIMS analysis technique using element standards to precisely quantify the La substitution concentration as well as to detect impurity elements from the high-temperature solutions in our sub-micrometer LPE films. Furthermore, angular dependent measurements of the resonance field and of the FMR linewidth will be intended to determine the influence of uniaxial magnetic anisotropies on the ferromagnetic resonance losses.

In conclusion, liquid phase epitaxy has the potential to provide sub-micrometer YIG films with outstanding crystalline and magnetic properties to meet the requirements for future magnon spintronics with ultra-low effective losses if a drastic miniaturization down to the nanometer scale is possible. First sub-100\,nm lateral sized structures have presently been prepared\cite{33} which could be the next step to LPE-based microscaled spintronic circuits. The development of YIG LPE films with thicknesses below 100\,nm is now in progress and remains a big challenge for the classical thick-film LPE technique.

\section{Methods}

\subsection{Sample fabrication}

YIG films were grown from PbO-B$_2$O$_3$ based high-temperature solutions resistively-heated in a platinum crucible at about $900^\circ$C using standard dipping LPE technique. During different growth runs nominally pure YIG films were grown on one-inch (111) gadolinium gallium garnet (GGG) substrates to check the reproducibility of the sub-micrometer liquid phase epitaxial growth. For La substituted films La$_2$O$_3$ was added to the already used high-temperature solution. To remove solution remnants from the sample surfaces the holder had to be stored in a hot acidic solution after room temperature cooling. Afterwards the reverse side layer was removed by mechanical polishing from the double-side grown samples. Chips of different sizes were prepared by a diamond wire saw and sample surfaces were cleaned using ethanol, distilled water and acetone. The LPE film thickness was determined by X-ray reflectometry using a PANanalytical/X-Pert Pro system.

\subsection{Microstructural properties}

The root-mean-square surface roughness was determined by AFM measurements for each sample at three different regions over 25\,$\mu m^2$ ranges using a Park Scientific Instruments, M5. HR-XRD studies were performed by a five-crystal diffraction spectrometer of Seifert (3003 PTS HR) equipped with a four-fold Ge 440 asymmetric monochromator using CuK$_\alpha$ radiation. The resolution limit was $1\times 10^{-4}$\,deg. GGG substrate lattice parameters were obtained by the Bond method. Depth profile analyses were carried out by an Axis Ultra$^\mathrm{DLD}$ XPS system (Kratos Analytical Ltd.) using a mono-atomic argon-ion etching technique. Qualitative SIMS (Hiden Analytical) measurements were carried out. Here, a film area of $500 \times 500\,\mu m^2$ is irradiated by 5\,keV oxygen ions.

\subsection{Magnetic properties}

The vibrating sample magnetometer (Micro\-Sense LLC, EZ-9) was used to register the in-plane hysteresis loops of the YIG/GGG samples at room temperature. The external magnetic field $H$ was controlled with an error of $\leq$0.01\,Oe. To estimate the magnetization of the YIG films we removed a contribution of the GGG substrates from the total VSM signal. To monitor the in-plane anisotropy as a function of the crystallographic orientation, the hysteresis loops at the azimuthal angles $0^\circ\leq \varphi\leq 360^\circ$ were measured with an angular step of $3^\circ$. The FMR absorption spectra were registered with a vector network analyzer (Rohde \& Schwarz GmbH, ZVA 67) attached to a broadband strip\-line. The sample was disposed face-down over a strip\-line and the transmission signals ($S_{21}$ \& $S_{12}$) were recorded. During the measurements, a frequency of microwave signals with the input power of $-10$\,dBm (0.1\,mW) was swept across the resonance frequency, while the in-plane magnetic field $H$ was constant and measured with an accuracy of 1\,Oe. Each recorded spectrum was fitted by the Lorentz function and allowed us to define the resonance frequency and the FMR linewidth $\Delta H_\mathrm{FWHM}$ corresponding to the applied field $H$.

\bibliography{main12}

\section{Acknowledgements}
We acknowledge the partial financial support by Deutsche Forschungs\-gemeinschaft (DU 1427/2-1). We thank M.~Frigge for EPMA analysis, Ch.~Schmidt for XRR measurements and R.~Meyer and B.~Wenzel for technical support.

\section{Author contributions statement}
C.D. conceived the experiments, prepared all samples and analyzed the data. O.S. performed VSM and FMR measurements and analyzed the data. R.L. performed the XPS experiments. J.D. and U.B. performed the SIMS experiments. A.D. conducted the XRD experiments and analyzed the data. C.D. and O.S. wrote the manuscript. All authors contributed to scientific discussions and the manuscript review.

\section{Additional information}

\subsection{Competing financial interests}

The authors declare no competing financial interests. 

\end{document}